\documentclass{iau}
\usepackage{graphicx}

\newcommand{\apj}{{\it ApJ}}
\newcommand{\aj}{{\it AJ}}
\newcommand{\mnras}{{\it MNRAS}}
\newcommand{\aanda}{{\it A\&A}}

\title[IAUS289.~~AGB Variables as Distance Indicators] 
{Asymptotic Giant Branch Variables as Extragalactic Distance Indicators}

\author[Patricia A. Whitelock]   
{Patricia A. Whitelock$^{1,2}$}

\affiliation{$^1$ South African Astronomical Observatory, P. O. Box
9, 7935 Observatory, South Africa\\
$^2$ Astronomy, Cosmology and Gravity Centre, Astronomy Department,
 University of Cape Town, 7701 Rondebosch, South Africa\\}
   
\pubyear{2012}
\volume{289}  
\pagerange{1--8}
\jname{Advancing the Physics of Cosmic Distances}
\editors{Richard de Grijs \& Giuseppe Bono, eds.}
\begin{document}

\maketitle

\begin{abstract}
Large-amplitude asymptotic giant branch variables potentially rival
Cepheid variables as fundamental calibrators of the distance scale,
particularly if observations are made in the infrared, or where there
is substantial interstellar obscuration. They are particularly useful
for probing somewhat older populations, such as those found in dwarf
spheroidal galaxies, elliptical galaxies or in the halos of spirals.
Calibration data from the Galaxy and new observations of various Local
Group galaxies are described and the outlook for the future, with a
calibration from {\sl Gaia} and observations from the next generation
of infrared telescopes, is discussed.

\end{abstract}

\firstsection 
\section{Introduction}

The focus of this paper is on the large-amplitude, or Mira, asymptotic
giant branch (AGB) variables. Because of their high luminosity, and
particularly their high {\it infrared} luminosity, stars in this group
have huge potential as extragalactic distance indicators. I briefly
describe what we know about the Mira period--luminosity (PL)
relationship, before going on to review recent work on Miras in Local
Group galaxies, contrasting those found in dwarf spheroidals with
those in the dwarf irregular galaxy NGC\,6822.

Miras are very large-amplitude ($\Delta V > 2.5$, $\Delta K > 0.4$
mag), long-period ($P>100$ days) variables.  They are close to the
maximum bolometric luminosity that they will ever achieve and they are
cool, with spectra dominated by molecular absorption.  Stars with
initial masses in the range from 0.8 to 8 M$_\odot$, possibly higher,
are thought to go through the Mira evolutionary phase.  For most of
their AGB lifetime, their nuclear energy comes from burning of the
hydrogen shell. However, from time to time stars on the upper AGB
experience helium-shell flashes with far-reaching consequences for the
stellar atmospheres.

It is following these flashes that dredge-up can occur, bringing
material from the core into the convective regime, and the atmospheric
abundance can change from oxygen (O)- to carbon (C)-rich.  In
relatively high-mass AGB stars, hot-bottom burning can also affect
abundances, turning carbon to nitrogen and producing lithium, among
other things.  The relative fraction of O- and C-rich Miras in any
population is a function of both age and metallicity.

\section{Period--Luminosity Relations} 

Early work on the Mira PL relation (e.g., at $K$ or $M_{\rm bol}$) in
the Large Magellanic Cloud (LMC) was published by Feast et al. (1989)
and Hughes et al.  (1990).  However, our understanding of this was
greatly advanced when Wood (2000) demonstrated that most AGB and
upper-giant-branch variables followed various parallel PL sequences.
The Mira PL relation, which had been known for some while, was just
one of them.  Wood showed that the Mira sequence corresponds to
fundamental pulsation and includes most of the large-amplitude
variables as well as a few of the low-amplitude, or semi-regular,
variables.  Some of the O-rich large-amplitude, long-period ($P>400$
days) Miras lay above the fundamental sequence, as had also been known
for some while.

The early studies of the multiple sequences used periods determined
from microlensing studies, e.g., from {\sc macho} and {\sc ogle} at
$V$ and $I$, and these results are now being refined as more
systematic observations are made at infrared wavelengths.  Among other
things, the situation is obviously confused by the fact that many of
the Miras have circumstellar shells.  In some cases these are so thick
that they affect the position of the star on the PL($K$) relation
(e.g., Ita \& Matsunaga 2011).  In those cases, it is preferable to
work with bolometric magnitudes, if they can be reliably estimated,
although it is sometimes possible to correct for circumstellar
extinction.

The Miras which lie above the PL relation are an interesting subgroup
that may be experiencing hot-bottom burning; there is certainly
evidence for lithium and other enrichment (Whitelock et al.  2003;
Whitelock 2012; and references therein).  It also seems possible that
they are pulsating in the first overtone (Feast 2009).

From the distance-scale perspective, the multiple PL sequences of the
small-amplitude variables are not very useful, since it is unclear
which sequence any particular star will be on. It is therefore only
the Miras, i.e., the large-amplitude variables, that we consider
useful for distance-scale studies and, where possible, Miras with
short periods ($P<400$ days) and thin dust shells are easier to work
with.

Whitelock et al. (2008) discuss the Galactic calibration of the
PL($K$) relation and show that it is consistent with the LMC relation.
Rejkuba (2004) demonstrated that the PL($K$) relation for O-rich Miras
in the LMC also fitted similar stars in NGC\,5128.  This, together
with recent work on Miras in dwarf spheroidals and NGC\,6822 (see
below) is consistent with the same PL relation applying everywhere.
It remains possible that there are metallicity effects, but these are
unlikely to be significantly greater than 0.1 mag for stars from
populations considered so far (see Matsunaga 2012).

\begin{figure}
\begin{center}
 \includegraphics[width=3.4in]{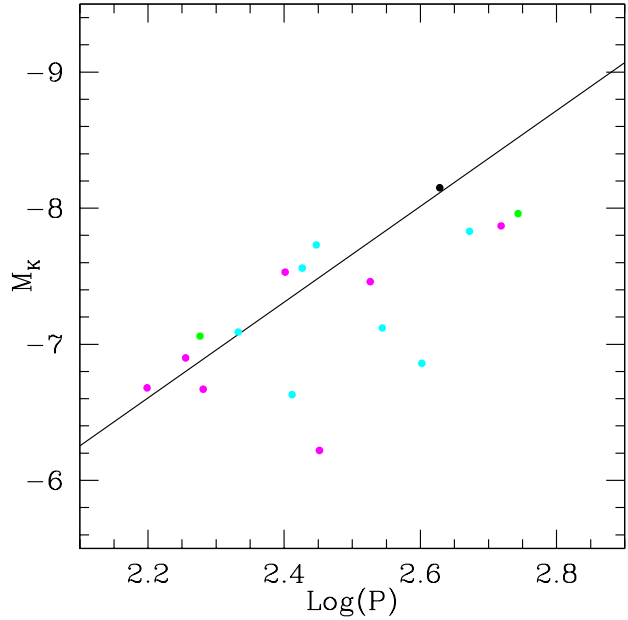}
 \caption{PL($K$) relation for the Miras in the dwarf spheroidals,
   coded as follows: Fornax (cyan), Leo~{\sc i} (magenta), Sculptor
   (green), Phoenix (black).  The line is at the slope of the LMC PL
   relation and assumes a distance modulus to the LMC of 18.39 mag
   (changing this distance modulus will not alter the loci of the
   points and the line, only their place relative to the magnitude
   scale).}
   \label{fig1}
\end{center}
\end{figure}

\section{Which Mira PL relation for distance-scale studies?}

For most purposes at the present time it is preferable to work with
the PL($K$) relation, primarily because the $K$ magnitude is easy to
measure, the pulsation amplitude at $K$ is lower than at shorter
wavelengths and $K$ is not often severely affected by circumstellar
emission or absorption.  However, pulsation drives mass loss, and
there are Miras with thick shells and severe circumstellar reddening.
This is particularly true of the very long-period O-rich Miras, known
as OH/IR stars, but it is also the case for a variety of C-rich stars.

In principle using bolometric magnitudes avoids the problem with
circumstellar extinction and there are three related ways of deriving
such magnitudes, all of which have drawbacks:

\begin{enumerate}
\item Ideally, one obtains bolometric magnitudes derived from
  multi-epoch measurements across a wide wavelength range extending at
  least into the mid-infrared regime. Such data are rarely available for
  many stars.

\item An alternative approach uses well-calibrated colour-dependent
  (e.g., $J-K$) bolometric corrections, derived for stars that are
  similar to those of interest, and for which multi-wavelength
  observations are available.  A potential problem is that the
  observed colour is a combination of the intrinsic colour of the star
  and the reddening.  The same $(J-K)$ colour can be due to different
  combinations of these two effects.  Thus, $(J-K)$ may not
  necessarily be uniquely related to the true bolometric
  correction.

\item It is sometimes possible to make a correction for the
  circumstellar reddening (e.g., Matsunaga et al.  2009), which is
  similar to applying a colour-dependent bolometric correction.  The
  weakness of this is that it requires adoption of an intrinsic colour
  for the star and we have little evidence of the way this colour
  correlates with period, particularly for long-period stars, almost
  all of which have shells.
\end{enumerate}

Of course, non-uniform shells, which are the obvious consequence of
non-uniform mass loss, present obvious problems to any method for
estimating the bolometric magnitude. Different methods of calculating
the bolometric magnitude give very significantly different results.
For example, Groenewegen et al.  (2007) and Kamath et al.  (2010)
derive bolometric magnitudes for several pulsating stars in the Small
Magellanic Cloud (SMC) cluster NGC\,419 using slightly different
$JHKL$ values but the same {\sl Spitzer} data.  Their bolometric
magnitudes differ by amounts that range from $-0.1$ to 0.4 mag for the
same star (see also Kerschbaum et al. 2010).

\begin{figure}
\begin{center}
\includegraphics[width=3.4in]{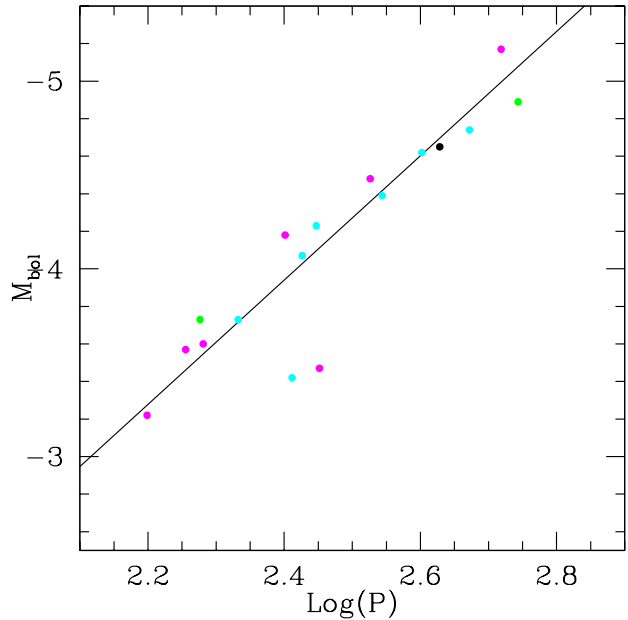}
 \caption{Bolometric PL relation for the Miras in the dwarf spheroidal
   galaxies, colours as in Fig.~1.}
   \label{fig2}
\end{center}
\end{figure}

Noting the challenge of determining accurate bolometric magnitudes,
estimated values can still provide useful distances, provided a very
systematic approach is followed.  In practice, this means determining
the bolometric magnitude of the star for which the distance is
required in {\it exactly} the same way as for the calibrators used to
define the PL relation.

\section{Globular Cluster: Lyng{\aa} 7}

Matsunaga (2006) discovered a Mira variable, V1, in Lyng{\aa}~7 (an
old, metal-rich Galactic bulge cluster) with a period of 551 days, a
large amplitude, $\Delta K = 1.22$ mag, and red colour, $(J-K)=4.1$
mag.  Sloan et al.  (2010) showed, based on a {\sl Spitzer} spectrum,
that it is carbon-rich and Feast et al.  (2012a) used a spectrum from
the Southern African Large Telescope ({\sl SALT}) to demonstrate that
V1 is a radial-velocity member of the cluster.

Assuming a distance modulus of 14.55 mag (Sarajedini et al. 2007),
this Mira has $M_{\rm bol}=-5.0$ mag, in agreement with the
PL-relation value of $M_{\rm bol}=-5.2$ mag.  Such a luminous star
must have had an initial mass $M_{\rm i} \sim 1.5 {\rm M}_{\odot}$ and
cannot be a normal member of the cluster.  It therefore must have
formed from a stellar merger.

To the best of our knowledge, this is the first ever demonstration of
a star that was produced by the merger of two others, but that
nevertheless obeys a PL relation. It is an interesting result and we
might well expect there to be remnants of other mergers in the dense
environment of the Galactic bulge.

\section{Local Group Galaxies}

A group of us from South Africa and Japan have used the Infrared
Survey Facility ({\sl IRSF}) at the South African Astronomical
Observatory to survey a variety of Local Group galaxies for AGB
variables.  The several-year survey uses observations made with the
{\sl SIRIUS} camera, which simultaneously gives $J$, $H$ and $K_{\rm
  s}$ photometry over a $7 \times 7$ arcmin$^2$ field.

\subsection{Dwarf Spheroidal Galaxies}

Results so far have been published for a total of 17 Miras from
Fornax, Leo~{\sc i}, Sculptor and Phoenix (Menzies et al. 2008, 2010,
2011; Whitelock et al. 2009). Where spectral types are available for
these stars, they show them to be C-rich and we assume that they are
all C-type stars.
 
Fig. 1 shows the absolute $K$ magnitudes on a PL($K$) relation for all
the Miras in dwarf spheroidals. The large scatter is very striking and
the distance below the LMC's PL($K$) relation is a function of the
$(J-K)$ colour, indicating that the stars below the line are there
because of their thick circumstellar shells.

Bolometric magnitudes can be estimated using a $(J-K)$-dependent
bolometric correction (Whitelock et al.  2009) and the results are
shown in a PL relation in Fig.~2.  The scatter is vastly reduced,
although there are still two stars which lie well below the mean
relation.  Two possible explanations have been offered for these faint
points.  It may be that the bolometric correction does not apply to
these intrinsically relatively blue, short-period, stars (see point
[{\it b}] in Sect. 3), since it was derived for significantly
longer-period stars.  Alternatively, they are undergoing obscuration
events of the type that are common among C-rich Miras in the Galaxy
and the LMC (e.g., Whitelock et al.  2006) and have non-uniform
shells.

\begin{figure}
\begin{center}
 \includegraphics[width=5in]{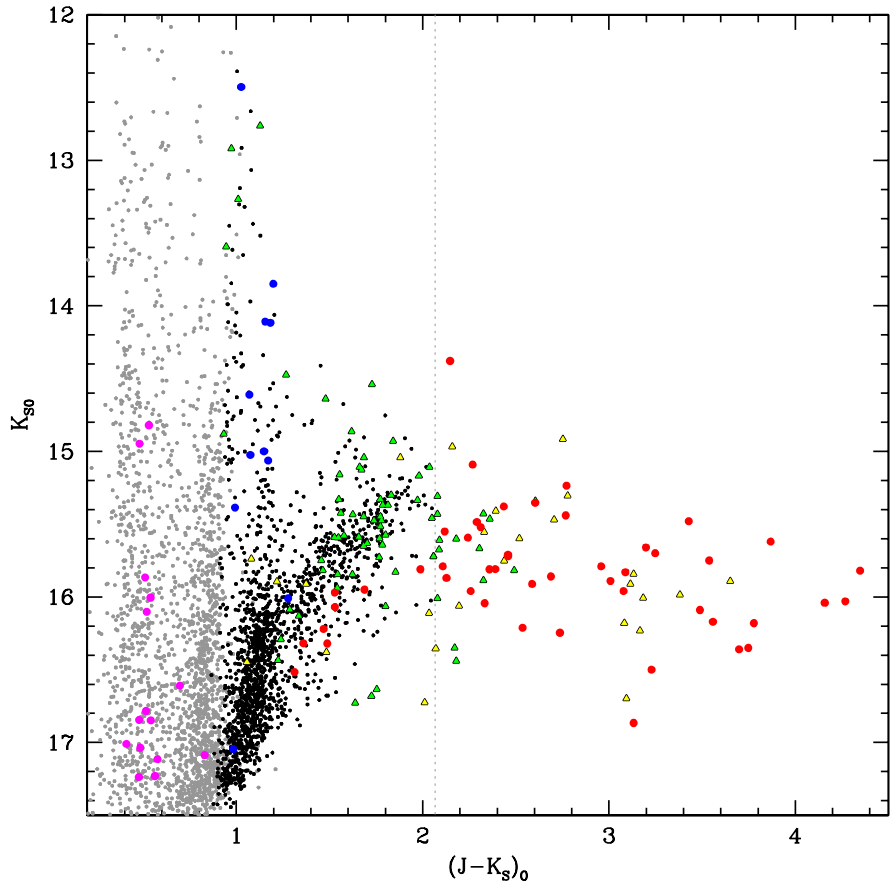}
 \caption{Colour--magnitude diagram for the variables in NGC\,6822,
   showing Cepheids (magenta), large-amplitude variables with measured
   periods, which are presumed O-rich (blue) or C-rich (red), and
   variables without measured periods that have large (yellow) or
   small (green) amplitudes.}
   \label{fig3}
\end{center}
\end{figure}

\subsection{NGC\,6822}

NGC\,6822 is an isolated barred dwarf galaxy, similar to the SMC, but
with slightly higher metallicity.  It has been examined for AGB
variables in the same way as the dwarf spheroidals over 3.5 years and
has had numerous Miras catalogued (Whitelock et al.  2012; see also
Battinelli \& Demers 2011).  Fig.~3 shows the variables in a
colour--magnitude diagram.  Spectral types are available for only very
few stars and the split into O- and C-rich assumes that all very red
stars are C-rich.  Several of the large-amplitude stars without
measured periods are probably also Miras and all very red stars are
variable.

\begin{figure}
\begin{center}
 \includegraphics[width=3.4in]{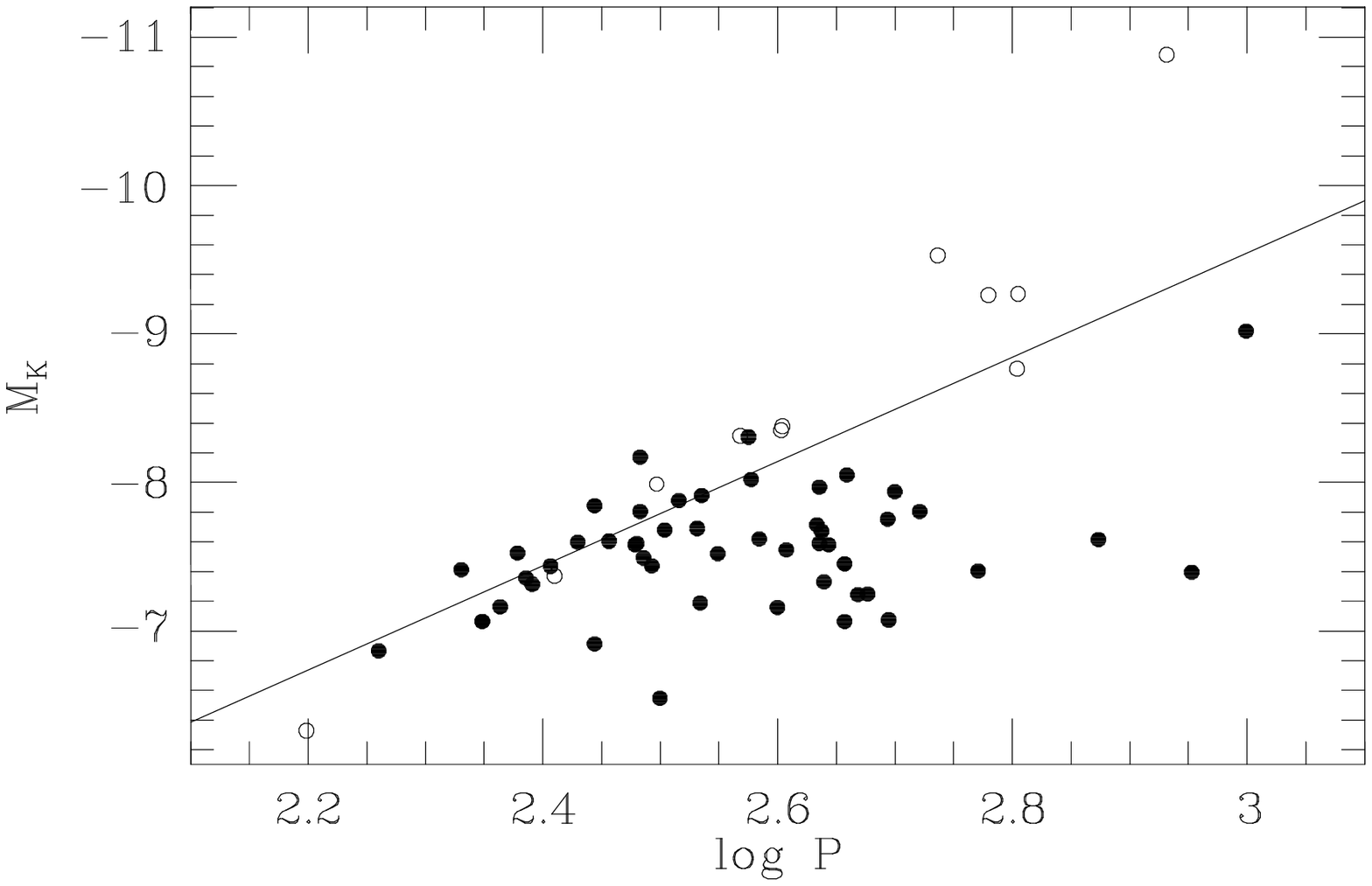}
 \caption{PL($K$) relation for NGC\,6822 showing O-rich (open circles)
   and C-rich Miras.  The line has the slope of the LMC PL($K$)
   relation and has been fit to the four short-period O-rich Miras.  }
   \label{fig5}
\end{center}
\end{figure}

Fig. 4 shows the NGC\,6822 Miras on a PL($K$) relation. Most of the
longer-period O-rich stars fall above the PL relation and are probably
similar to the stars in the LMC (mentioned above) that may be
hot-bottom burning (Whitelock et al.  2003). Feast (2009) suggested
that these stars may be pulsating in the first overtone. Many of the
C-rich Miras fall well below the line as the result of thick
circumstellar shells. Fig.~5 shows the same stars on a bolometric PL
relation, and we see that the C stars scatter around a relation that
is very similar to the one obeyed by LMC Miras. The slope is almost
identical to the slope of the LMC line, within the uncertainties.

Using an LMC distance modulus of 18.5 mag, we determine from the
C-rich Miras that $(m-M)_0=23.56\pm 0.03$ mag for NGC\,6822.  This may
be compared to $23.40\pm 0.05$ mag derived from Cepheid variables
(Feast et al.  2012b) and $23.49\pm 0.03$ mag from RR Lyrae variables
(Clementini et al.  2003).  Note that all errors quoted here are
internal, but there are systematic uncertainties in all of the
measurements.  The agreement is reasonable and certainly shows that
Miras offer a viable alternative to the more conventional distance
indicators.

\subsection{Challenges}

However, I should note that there remain challenges in using Miras as
distance indicators.  One of the most serious of these is ensuring
that measurements made with different photometric systems give the
same result.  Battinelli \& Demers (2011) have 16 large-amplitude
variables in common with Whitelock et al.  (2012) in NGC\,6822.  The
periods determined for these agree well, but the mean magnitudes
differ by $\Delta K=0.25$ mag.  Both groups know that their photometry
of normal stars, i.e., those with $(J-K)<1.0$ mag, is on the {\sc
  2mass} system.  Dealing with very red stars, of which there are no
non-variable examples, will require considerably more effort.

It also remains possible that we will find metallicity effects as the
PL relationships become better defined.

\begin{figure}
\begin{center}
 \includegraphics[width=3.4in]{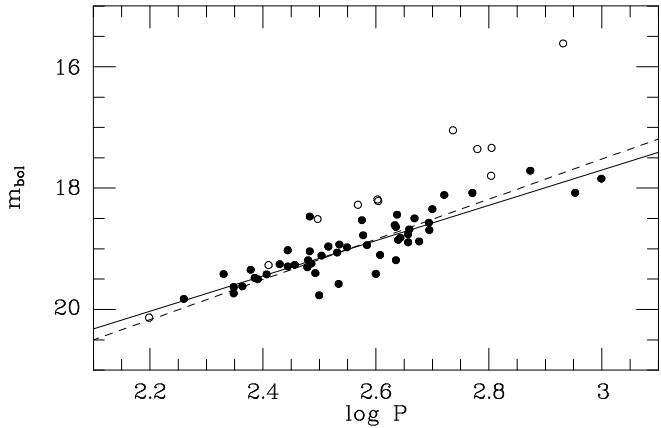}
 \caption{Bolometric PL relation for NGC\,6822, showing the same stars
   as in Fig.~5. The dashed line is the LMC PL relation, while the
   solid line is the relation fitted to these data.}
   \label{fig6}
\end{center}
\end{figure}

\section{Cepheids and/or Miras?}

Cepheids have long provided a vital step on the distance-scale ladder
linking the Galaxy to distant supernovae. However, Mira variables
offer a viable alternative which may well be preferable for the
following reasons:
\begin{itemize}
\item Miras are of comparable brightness to Cepheids at $K$ and
  brighter at longer wavelengths (see Table~1 and Whitelock 2012).
\item Miras are found in galaxies which do not host Cepheids, such as
  dwarf spheroidals, and will be found in ellipticals.
\item Miras are found in the haloes of spiral galaxies, where they may
  be less confused, and therefore more easily observable at large
  distances, than stars in the spiral arms.
\item Miras are best observed in the infrared and are therefore not
  severely affected by interstellar extinction.
\item The next generation of space telescopes, as well as large
  ground-based telescopes equipped with adaptive optics, will
  primarily work in the infrared. They will be ideally suited to
  observing distant Mira variables.
\end{itemize}

 \begin{table}
\begin{center}
 \caption{Comparison of the absolute magnitudes of Cepheid and Mira
   variables at 2.2 and 8~$\mu$m (Whitelock 2012 and Feast 2010,
   unpublished).}
 \label{tbl}
 \begin{tabular}{cccc}
 \hline  
Wavelength & Variable & Period & Absolute \\
           &  type   & (days)  & mag \\
\hline
$K$(2.2 $\mu$m) & Cep & 50 & $-7.9$\\
               & M   & 380 & $-7.9$\\
\hline
  8 $\mu$m      & Cep & 50 & $-8.3$\\
               & M   & 230 & $-8.3$\\
               & M   & 380 & $-9.2$\\
 \hline 
 \end{tabular}
\end{center}  
 \end{table}  
 
\section{Conclusion}

Large-amplitude AGB variables offer a viable alternative to Cepheids
for distance-scale studies, which will be particularly valuable when
infrared observations are available. There remain, however,
calibration issues that must be resolved if observations from
different instruments are to be combined reliably. The {\sl Gaia}
satellite will provide a vital Galactic calibration that will put the
Mira absolute-magnitude scale on a new footing (see Whitelock 2012).

\acknowledgments I am grateful to my colleagues for allowing me to
discuss our results and particularly to Michael Feast and John Menzies
for their comments on this manuscript. I acknowledge a grant from the
South African National Research Foundation.

\end{document}